\documentclass{article}
\usepackage{spconf,amsmath,epsfig}

\usepackage[]{graphicx}
\graphicspath{{Figures/}}
\usepackage{caption}
\usepackage{subfig}
\usepackage{amsmath}
\usepackage{amssymb}
\usepackage{epsfig}
\usepackage{cite}
\usepackage{color}
\usepackage{balance}
\usepackage{amsmath}
\usepackage{amsfonts} 
\usepackage{graphicx}
\usepackage{pdfpages}
\usepackage[T1]{fontenc} 
\usepackage{array}
\usepackage{url}

\usepackage{color}
\usepackage{wrapfig}
\usepackage{lipsum}
\usepackage[hidelinks]{hyperref}
\usepackage{multirow}
\usepackage{tabularx}
\usepackage{gensymb} 
\usepackage{breqn}
\usepackage{array}
\usepackage{booktabs}

\title{An Ontology Model for Climatic Data Analysis}

\name{Jiantao Wu$^{1,2}$, Fabrizio Orlandi$^{1,3}$, Declan O'Sullivan$^{1,3}$, and Soumyabrata Dev$^{1,2}$
\thanks{This research was partially funded by the EU H2020 research and innovation programme under the Marie Skłodowska-Curie Grant Agreement No.~713567 at the ADAPT SFI Research Centre at Trinity College Dublin. The ADAPT Centre for Digital Content Technology is funded under the SFI Research Centres Programme (Grant 13/RC/2106) and is co-funded under the European Regional Development Fund.
}
\thanks{Send correspondence to S.\ Dev: \url{soumyabrata.dev@ucd.ie}}
}
\address{
	$^{1}$~The ADAPT SFI Research Centre, Ireland \\
	$^{2}$~School of Computer Science, University College Dublin, Ireland \\     
    $^{3}$~School of Computer Science and Statistics, Trinity College Dublin, Ireland
}

\begin{document}
%\ninept
%
\maketitle
\begin{abstract}
Recently ontologies have been exploited in a wide range of research areas for data modeling and data management. They greatly assists in defining the semantic model of the underlying data combined with domain knowledge. In this paper, we propose the Climate Analysis (CA) Ontology to model climate datasets used by remote sensing analysts. We use the data published by National Oceanic and Atmospheric Administration (NOAA) to further explore how ontology modeling can be used to facilitate the field of climatic data processing. The idea of this work is to convert relational climate data to the Resource Description Framework (RDF) data model, so that it can be stored in a graph database and easily accessed through the Web as Linked Data. Typically, this provides climate researchers, who are interested in datasets such as NOAA, with the potential of enriching and interlinking with other databases. As a result, our approach facilitates data integration and analysis of diverse climatic data sources and allows researchers to interrogate these sources directly on the Web using the standard SPARQL query language. 
\end{abstract}

\section{Introduction}
\label{sec:intro}
A large number of today's climate data centers present their collected data in the form of raw tables (\textit{e.g.} RDB, CSV, JSON): KNMI Climate Explorer\footnote{\url{http://climexp.knmi.nl/}}, NOAA datasets\footnote{\url{https://www.ncdc.noaa.gov/data-access},\label{noaa}}. These datasets are helpful when people are researching a limited number of climatic factors in a single data source, \textit{e.g.} analyzing how temperature changes over time in a certain city. As this might only need to interrogate a few number of columns within a data table. However, climatic changes can be commonly influenced by some other factors beyond the climate domain, such as urbanization~\cite{chu2016temperature}. Including data of different domains can be a very time-consuming job for climatic researchers. Recently, one of the popular solutions that is greatly explored is employing an ontology\footnote{See Section~\ref{ontologies} for a detailed definition.}, that offers the expressivity and flexibility to easily extend to various interoperable domains~\cite{Orlandi2019}. In this paper, we propose CA ontology which reuses open SOSA ontology as a base for a general description of sensor networks presented in NOAA datasets. It also specifies some of generic terms in SOSA, which can be directly used to describe some complex climate features in the dataset and thus bring more convenient query experience (\textit{e.g.} reducing complexity of query patterns). Technically, it extends some of the classes and attributes in SOSA with additional sub-classes of climatic features (\textit{e.g.} `temperature', `precipitation', `wind speed'). For instance, in the CA ontology `TemperatureObservation' is defined a sub-class of SOSA `Observation' and CA `TemperatureSensor' is a sub-class of SOSA `Sensor'. Next, we publish the NOAA climate dataset as Linked Data (see Section~\ref{section:linked data} for more details) on the Web, which is available for being further exploited by other linked datasets. In addition, from the perspective of linked data characteristics, it also has the potential of serving as a knowledge graph (KG)~\cite{bonatti_et_al:DR:2019:10328}, for instance, allowing for knowledge inference if rules are defined~\cite{hogan2020knowledge}. However, KG inference is not explored in this paper.

The rest of the paper\footnote{\label{note3}In the spirit of reproducible research, all the source code is available at \url{https://github.com/futaoo/codespaceRepo}.} will be organized as follows: Section~\ref{section:related work} summarizes the related work and to what extent they are adopted in this paper, including the description of the chosen datasets, the ontology concepts, Linked Data and our proposed CA ontology. Section~\ref{section:data modeling} details how CA is defined and the mapping process from raw table data to RDF triples and then Linked Data publication. Section~\ref{section: a sample analsis} offers a real use case of collecting data for climatic analysis so as to show the convenience of using SPARQL query language to obtain experimental data. Finally, conclusions about the current phase of the work and our future work will be given in Section~\ref{section: conclusion}. 

\section{Background}
\label{section:related work}
\subsection{NOAA Climate Dataset}

The NOAA Climate Data Online service (CDO\footnote{\url{https://www.ncdc.noaa.gov/cdo-web/}}), is the primary raw data source used as a driving use-case to populate the proposed ontology in this paper. CDO covers some main climate features including precipitation, temperature, sea level pressure, wind speeds and so forth. These data are recorded in different time intervals (\textit{e.g.} daily summaries, precipitation hourly), which indeed provides great flexibility to set an appropriate time unit for climate analysis. 

Importantly, NOAA's diversiform data in climate domain proffers us convincing evident such that potential climate researchers can broaden their research horizons in the future with the integration of the proposed platform as one of powerful analysis medium based on linked data and knowledge graph principles.

\subsection{Ontologies in short}
\label{ontologies}
This section will give a high-level definition of ontology followed by some open fundamental ontologies employed for building the CA ontology in this paper.

According to W3C's definition\footnote{\url{https://www.w3.org/standards/semanticweb/ontology}}: `ontologies' ( or `vocabularies' in the simple term) unambiguously define concepts and relationships for specific fields of concern. Take wireless sensor network as example, ontology can be made to classify different types of sensors by simply naming `wind sensor', `temperature sensor', etc. And the corresponding observations can be connected with by creating links named, for example, `hasResults'. Consequently with a set of terms defined, a semantic layer is formed consisting of operative human describable terms over the data. Here we list some main open ontologies and vocabularies in the climate domain used for our purpose.

\textbf{SOSA Ontology\footnote{\url{https://www.w3.org/TR/2017/REC-vocab-ssn-20171019/}}}. This ontology is applied in CA to representation of sensor related vocabularies such as sensors, observations, samples, etc.

\textbf{Climate and Forecast (CF) vocabulary\footnote{\url{https://www.w3.org/2005/Incubator/ssn/ssnx/cf/cf-property}}}. All of the geophysical features (\textit{e.g.} `air\_temperature') in this paper are defined in line with CF. Terminologies in CF can be parsed through a certain syntax for meaning, which is explained at the documentation \footnote{\url{http://www.met.reading.ac.uk/~jonathan/CF_metadata/14.1/}}.

In addition, this paper also refers to some other complementary ontologies, namely `wgs84' ontology\footnote{\url{https://www.w3.org/2003/01/geo/}} for geographical locations, `qudt' units vocabulary\footnote{\url{http://www.qudt.org/}} for units of sensor observed results.

\subsection{Linked Data in short}
\label{section:linked data}

Linked Data is a research field and a set of principles that deal with how data is structured, interconnected and published on the Web~\cite{bizer2011linked}. It allows people to navigate data through the data links from one data source to another, which enables a fixed set of data sources to be easily extended with all of the other linked data sources available on the Web~\cite{Radulovic2015}. One of the advantages brought by this behavior is such that --- if needed --- federated queries (or reasoning) can be performed on such global wide database on the Web to gain resources (or knowledge) beyond the fixed databases~\cite{8706177}. 

To leverage the Linked Data principles, we choose Apache Jena Fuseki\footnote{\url{https://jena.apache.org/documentation/fuseki2/}} as storage layer and also as publishing tool. Fuseki is an open source SPARQL server with functionality in storing RDF triples and performing RDF queries. It supports SPARQL 1.1 Graph Store HTTP Protocol\footnote{\url{https://www.w3.org/TR/sparql11-http-rdf-update/}}, \textit{i.e.} it allows users to query a Fuseki database via HTTP requests (\textit{e.g.} HTTP GET). At the same time, the RDF triplestore in Fuseki takes the role of the endpoint accessible to other remote endpoints. In turn, making it easier to link remote data source services via SPARQL queries.

\section{Data Modeling}
\label{section:data modeling}
\subsection{CA ontology}
\begin{figure}[ht]
\centering
\includegraphics[scale=0.35, trim = 0cm 18cm 0cm 0cm, clip]{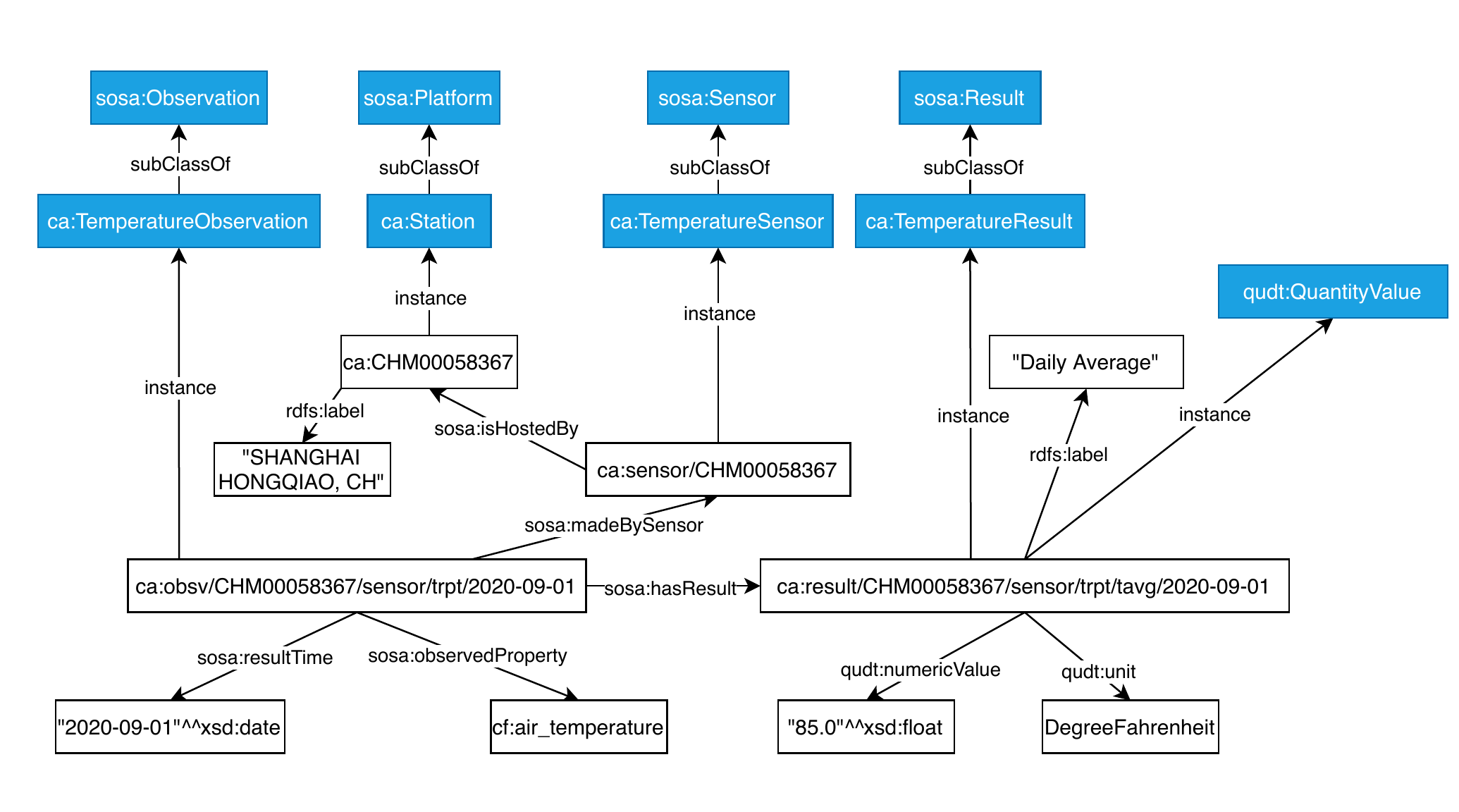}
\caption{The average temperature record of Shanghai on 2020-09-01 represented using our CA ontology}
\label{fig:ontology}
\end{figure}

In this section, we describe our CA ontology. CA ontology is built for Climate and weather data Analysis. As previously described (Sec.~\ref{section:related work}), it is built reusing and extending existing ontologies and provides a graph data schema for CDO's daily climatic sensor records. The diagram in Fig.\ref{fig:ontology} shows how a daily average temperature observation is represented using the CA ontology. 

The following lists some of the main classes and properties of concern in CA ontology. To note that, for brevity, the question mark `[?]' denotes that it is a specified climate factor as mentioned in section~\ref{sec:intro} like `Temperature'. It can be replaced with different climate features refining the more general super-class it is derived from.\footnote{Please refer to our GitHub repository for more details about the ontology: \url{https://github.com/futaoo/codespaceRepo}}

\textbf{ca:[?]Observation}. A subclass of sosa:Observation, which indicates an observation is about `?'.

\textbf{ca:Station}. A subclass of sosa:Platform. A station is a place for the setup of sensors.

\textbf{ca:[?]Sensor}. A subclass of sosa:Sensor, indicating a sensor for recording `?' data.

\textbf{ca:[?]Result}. A subclass of sosa:Result. Here the result must be related to `?'. It could be a data series for a period of time or a sole value at a certain timestamp.

\textbf{sosa:isHostedBy}. A property to describe the host of a sensor.

\textbf{sosa:hasResult}. A property to describe a result is produced by an observation.

\textbf{sosa:isMadeBySensor}. A property to describe an observation is made by a sensor.

\textbf{sosa:resultTime}. A property to describe the time when an observation is made.

\textbf{sosa:observedProperty}. A property to describe a specific quality of the feature of interest associated with an observation.

\subsection{Data conversion}
This procedure achieves the retrieval of the relational CDO dump data and storage of them as RDF triples in our Fuseki triplestore. We developed a Python data transformation pipeline and had published it on our Git repository. It exploits the Python package rdflib\footnote{\url{https://rdflib.readthedocs.io/en/stable/}} to accomplish the mapping from relational data to Turtle-syntax\footnote{\url{https://www.w3.org/TeamSubmission/turtle/}} based RDF triples which can be directly recognized by Fuseki SPARQL endpoint and then uploaded to it.

\section{Analyzing climate data using CA ontology}
\label{section: a sample analsis}

In this section, we show our main motivation for conducting this work through some sample use cases in regard to how our created CDO data SPARQL endpoint can be used in climate analytics.  

\subsection{Data collection}
\label{section: Data Collection}
CA ontology provides the means by which to query the climate database in the form of graph pattern~\cite{hogan2020knowledge}. Listing 1 is a sample SPARQL query that can be executed in our SPARQL endpoint. It was written in graph pattern for generating recent seven decades (1951-2020) of archives about daily precipitation and temperature conditions in Dublin and Shanghai.

\begin{figure}[htb]
\centering
\includegraphics[width=0.5\textwidth]{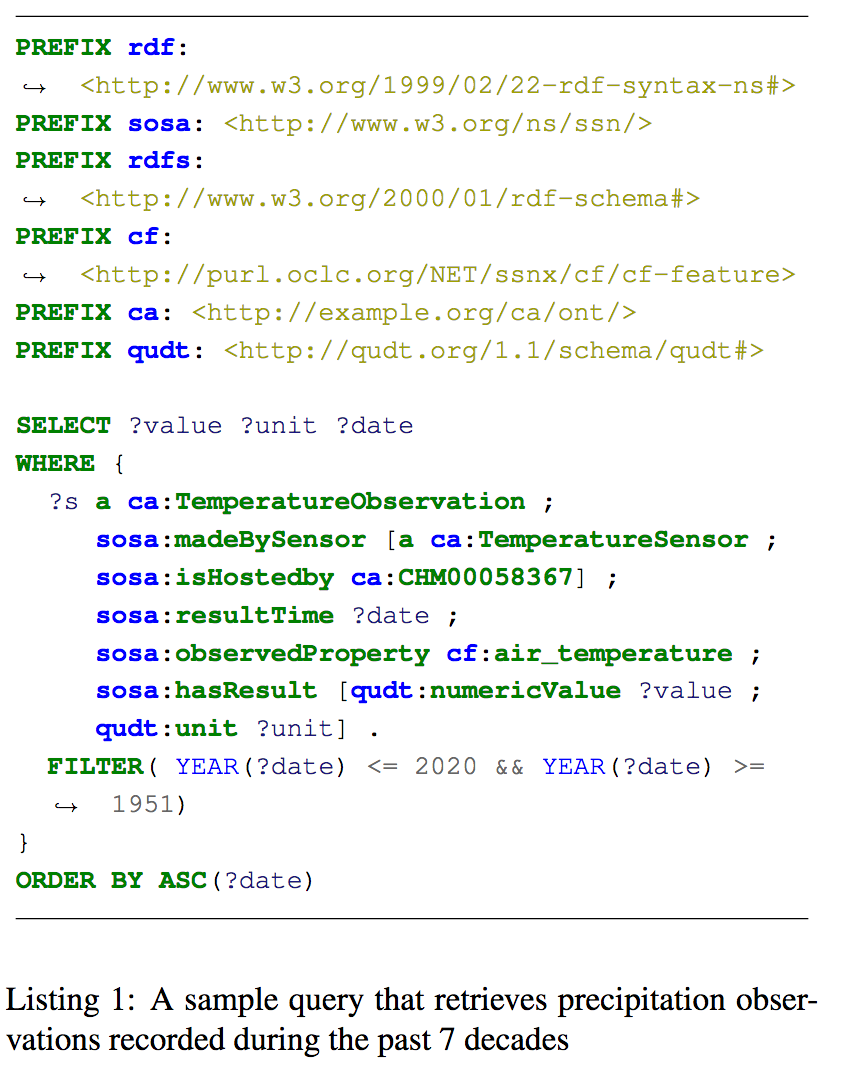}
\end{figure}

\vspace{-0.7cm}

\subsection{Obtaining automated weather summaries}
\begin{figure}[ht]
\centering
\includegraphics[scale=0.5]{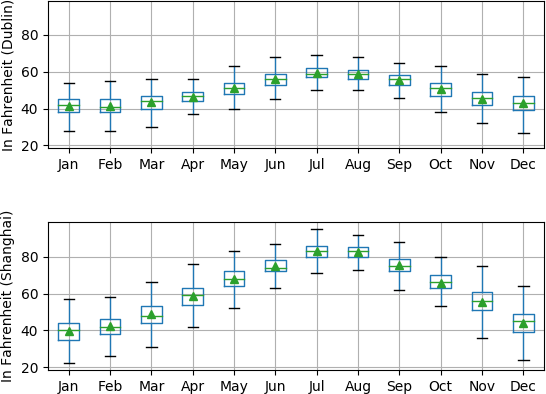}
\caption{Monthly temperature summarized from recent 70 years of Shanghai and Dublin}
\label{temperature}
\end{figure}

On the premise that someone is interested in the temperature difference between Shanghai and Dublin, a pair of box-plot (Fig.\ref{temperature}) can be drawn using previously yielded data (Section~\ref{section: Data Collection}), wherein each box includes monthly statistics of 70-year daily temperature records, and the triangle and green line segment denote the mean and the median respectively. Analyzing these figures at least following some weather summaries can be concluded.

\textbf{Temperature Commonalities and Differences}. A common characteristic of both the cities is daily temperature is very likely to be a \textbf{normal distribution} in a certain month. Besides, as time proceeds, the temperature will increase until reaching the highest in July and then decrease to the lowest in December. The only obvious difference is that Shanghai's temperature especially in summer is higher than Dublin.

\begin{figure}[htb]
\centering
\includegraphics[scale=0.5]{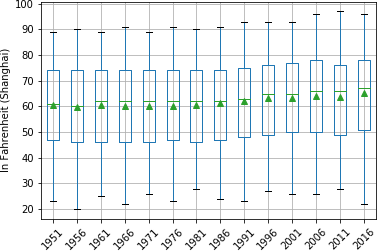}
\caption{The trend of temperature in Shanghai in last seven decades}
\label{fig: globalwarming}
\end{figure}

Additionally, it is easy to find a slightly rising trend of temperature in Shanghai since 1951 if plotting a diagram like Fig.\ref{fig: globalwarming}. In this figure, each coordinate value of axis x means a 5 years period starting from that value, for example, `1951' means a 5-year period from 1951 to 1955. This summary could be seen as an evidence of global warming impact on Shanghai which is exactly consistent with the conclusion given by Chu \textit{et al.} that Shanghai is subject to a serious temperature growth~\cite{chu2016temperature}.

\vspace{-0.5cm}

\vspace{-0.7cm}

\section{Conclusion \& Future Work}
\label{section: conclusion}
This paper proposes the CA ontology for modeling climate observations as Linked Data on the Web. We described our use-case which is focused on the transformation of NOAA / CDO data into RDF. To make CDO data compatible with the CA ontology, this work implements a mapping from raw CSV formatted data to RDF triples using the Python environment. Afterwards, the resulting RDF data is stored into a triplestore and published, using Jena Fuseki, so that it is allowed to be accessed as an endpoint on the Web. Lastly, we did a sample comparative climatic analysis between Dublin and Shanghai in precipitation and temperature to demonstrate how users could take full advantages of CDO data to acquire climatic insights conveniently through SPARQL queries combined with statistical analysis.

As part of our future work, we are aiming to improve the work on three aspects. First, there will be a continuous extension to the types of the classes and properties in CA ontology, catering for different datasets~\cite{dev2017nighttime}. Second, after having a rich set of vocabularies, define some climate domain-specific rules to bring the CA ontology an inference ability (\textit{i.e.} exploring the knowledge graph applications). Third, leverage the power of Linked Data to a certain degree by integrating CA ontology with other published ontologies in order to enable users to have some knowledge on how climate influences other domains (\textit{e.g.} a city's traffic).

\balance 



\begin{thebibliography}{1}
\providecommand{\url}[1]{#1}
\csname url@samestyle\endcsname
\providecommand{\newblock}{\relax}
\providecommand{\bibinfo}[2]{#2}
\providecommand{\BIBentrySTDinterwordspacing}{\spaceskip=0pt\relax}
\providecommand{\BIBentryALTinterwordstretchfactor}{4}
\providecommand{\BIBentryALTinterwordspacing}{\spaceskip=\fontdimen2\font plus
\BIBentryALTinterwordstretchfactor\fontdimen3\font minus
  \fontdimen4\font\relax}
\providecommand{\BIBforeignlanguage}[2]{{%
\expandafter\ifx\csname l@#1\endcsname\relax
\typeout{** WARNING: IEEEtran.bst: No hyphenation pattern has been}%
\typeout{** loaded for the language `#1'. Using the pattern for}%
\typeout{** the default language instead.}%
\else
\language=\csname l@#1\endcsname
\fi
#2}}
\providecommand{\BIBdecl}{\relax}
\BIBdecl

\bibitem{chu2016temperature}
W.~Chu, S.~Qiu, and J.~Xu, ``Temperature change of shanghai and its response to
  global warming and urbanization,'' \emph{Atmosphere}, vol.~7, no.~9, p. 114,
  2016.

\bibitem{Orlandi2019}
\BIBentryALTinterwordspacing
F.~Orlandi, A.~Meehan, M.~Hossari, S.~Dev, D.~O'Sullivan, and T.~AlSkaif,
  ``Interlinking heterogeneous data for smart energy systems,'' \emph{CoRR},
  vol. abs/1907.02790, 2019. [Online]. Available:
  \url{http://arxiv.org/abs/1907.02790}
\BIBentrySTDinterwordspacing

\bibitem{bonatti_et_al:DR:2019:10328}
\BIBentryALTinterwordspacing
P.~A. Bonatti, S.~Decker, A.~Polleres, and V.~Presutti, ``{Knowledge Graphs:
  New Directions for Knowledge Representation on the Semantic Web (Dagstuhl
  Seminar 18371)},'' \emph{Dagstuhl Reports}, vol.~8, no.~9, pp. 29--111, 2019.
  [Online]. Available: \url{http://drops.dagstuhl.de/opus/volltexte/2019/10328}
\BIBentrySTDinterwordspacing

\bibitem{hogan2020knowledge}
A.~Hogan, E.~Blomqvist, M.~Cochez, C.~d'Amato, G.~de~Melo, C.~Gutierrez,
  J.~E.~L. Gayo, S.~Kirrane, S.~Neumaier, A.~Polleres \emph{et~al.},
  ``Knowledge graphs,'' \emph{arXiv preprint arXiv:2003.02320}, 2020.

\bibitem{bizer2011linked}
C.~Bizer, T.~Heath, and T.~Berners-Lee, ``Linked data: The story so far,'' in
  \emph{Semantic services, interoperability and web applications: emerging
  concepts}.\hskip 1em plus 0.5em minus 0.4em\relax IGI Global, 2011, pp.
  205--227.

\bibitem{Radulovic2015}
\BIBentryALTinterwordspacing
F.~Radulovic, M.~Poveda-Villal{\'{o}}n, D.~Vila-Suero,
  V.~Rodr{\'{i}}guez-Doncel, R.~Garc{\'{i}}a-Castro, and
  A.~G{\'{o}}mez-P{\'{e}}rez, ``{Guidelines for Linked Data generation and
  publication: An example in building energy consumption},'' \emph{Automation
  in Construction}, vol.~57, pp. 178--187, 2015. [Online]. Available:
  \url{http://dx.doi.org/10.1016/j.autcon.2015.04.002}
\BIBentrySTDinterwordspacing

\bibitem{8706177}
F.~{Orlandi}, J.~{Debattista}, I.~A. {Hassan}, C.~{Conran}, M.~{Latifi},
  M.~{Nicholson}, F.~A. {Salim}, D.~{Turner}, O.~{Conlan}, D.~{O'sullivan}, and
  J.~{Tang}, ``Leveraging knowledge graphs of movies and their content for
  web-scale analysis,'' in \emph{2018 14th International Conference on
  Signal-Image Technology Internet-Based Systems (SITIS)}, 2018, pp. 609--616.

\bibitem{dev2017nighttime}
S.~Dev, F.~M. Savoy, Y.~H. Lee, and S.~Winkler, ``Nighttime sky/cloud image
  segmentation,'' in \emph{2017 IEEE International Conference on Image
  Processing (ICIP)}.\hskip 1em plus 0.5em minus 0.4em\relax IEEE, 2017, pp.
  345--349.

\end{thebibliography}
\end{document}